\documentclass[prl,showpacs,twocolumn]{revtex4}

\usepackage{graphicx}
\usepackage{amsmath}
\usepackage{slashed}
\usepackage{feynmp}

\begin{document}
\title{Dynamical Origin of Seesaw}
\author{Ji\v r\'{\i} Ho\v sek}
\email{hosek@ujf.cas.cz}
\affiliation{Department of Theoretical
Physics, Nuclear Physics Institute, Czech Academy of Sciences, 25068
\v Re\v z (Prague), Czech Republic}

\begin{abstract}
In anomaly free gauged three-flavor $SU(3)_f \times SU(2)_L \times
U(1)_Y$ model of Yanagida with fermion and gauge boson masses
described by conveniently chosen elementary scalar Higgs fields the
neutrino mass matrix comes out in the seesaw form. Following
Yanagida's suggestion we demonstrate that no Higgs fields are
needed. {\it Strong flavor gluon interactions themselves}, treated
in a separable approximation, result in universally split lepton and
quark masses calculated in terms of a few parameters. While the
realistic splitting of charged lepton and quark masses requires the
electroweak and QCD radiative corrections the neutrino seesaw mass
matrix comes out exact.

\end{abstract}


\maketitle

\section{I. Introduction}
Understanding the measured wide and irregular charged lepton and
quark mass spectrum is a nightmare of theoretical elementary
particle physics. Understanding the extreme lightness of the
observed neutrinos is its almost unbearable stage. Description of
these phenomena, however, does exist. The charged lepton and quark
masses are described in the Standard model by the Higgs mechanism
\cite{higgs}, and the extremely light neutrinos are described in its
minimal extension by the seesaw \cite{seesaw}. It amounts to
postulating superheavy Majorana masses $M_{M}$ of three right-handed
sterile neutrinos $\nu_{R}$; the observed neutrino spectrum is given
by the diagonalized $6 \times 6$ symmetric matrix
\begin{equation}
\left(\begin{array}{ccc}0 & m_D\\
m_D^T & M_M\\
\end{array}
\right) \label{seesaw}
\end{equation}
where $m_{D}$ are three Dirac neutrino masses generated by the
ordinary Higgs mechanism. As a result the physical neutrino spectrum
consists of three superheavy Majorana neutrinos with masses $\sim
M_M$, and three active Majorana neutrinos with masses $m_{\nu} \sim
m_D^2/M_M$. Choosing the free parameters $m_D$ and $M_M$
appropriately we obtain the masses of three active (Majorana)
neutrinos at will.

This elegant phenomenological construction rises questions. First,
are there deeper reasons for postulating superheavy sterile
neutrinos ? If not, the very explanation of extreme lightness of the
observed neutrinos by postulating the existence of unobservably
heavy ones does not seem very deep. Second, why zero in the left
upper corner instead of the Majorana mass matrix $m_M$ of the
left-handed neutrinos of the Standard model ?

The existence of superheavy right-handed sterile neutrinos is
perfectly justified in the gauged three-flavor $SU(3)_f \times
SU(2)_L \times U(1)_Y$ model of T. Yanagida \cite{yanagida}. First,
the very existence of right-handed neutrinos is enforced by the
strong theoretical requirement of anomaly freedom. Second, their
massiveness is enforced by the experimental fact that the gauge
flavor $SU(3)_f$ symmetry is badly broken and yet unobserved. The
spontaneous breakdown of the gauge chiral $SU(3)_f \times SU(2)_L
\times U(1)_Y$ symmetry down to $U(1)_{em}$ is phenomenologically
described by the following weakly interacting condensing Higgs
fields: (1) The field $\chi^{ab}(6,1,0)$ gives rise to huge
different masses of all eight flavor gluons. Its Yukawa coupling
with sterile right-handed neutrinos and their charge conjugates
generates their huge Majorana masses. (2) The fields $\phi^a(8,2,1)$
and $\phi^0(1,2,1)$ are responsible mainly for the electroweak
symmetry breakdown. The values in parentheses are the representation
dimensions of $SU(3)_f$, $SU(2)_L$ and the values of the weak
hypercharge, respectively. The Majorana mass matrix $m_M$ of the
left-handed neutrinos stays zero because the elementary Higgs field
$\phi(6,3,-2)$ which would generate it \cite{seesaw2} was
deliberately not introduced.

In conclusion of \cite{yanagida} Yanagida notices that his model "is
a possible candidate for the spontaneous mass generation by
dynamical symmetry breaking \cite{njl}". Indeed, without scalar
sector the gauge $SU(3)_f \times SU(2)_L \times U(1)_Y$ model would
yield the fermion mass spectrum calculable in terms of a few free
parameters: (1) The coupling constant $h$ of the strongly coupled
$SU(3)_f$ or, due to the dimensional transmutation, its scale
$\Lambda$. (2) The coupling constants $g$ and $g'$ of the weakly
coupled $SU(2)_L$ and $U(1)_Y$, respectively. The numerical values
of weak hypercharges $Y_f$, $f=l_L, e_R,\nu_R,q_L,u_R,d_R$ of the
chiral electroweakly interacting fermion fields $f$ are not free
parameters. They are uniquely fixed by their electric charges
$Q=T_3+\tfrac{1}{2}Y$:
\begin{eqnarray*}
Y(l_L)&=&-1,\phantom{bbb}Y(e_R)=-2,\phantom{bbb}Y(\nu_R)=0\\
Y(q_L)&=&\tfrac{1}{3},\phantom{bbbb}Y(u_R)=\tfrac{4}{3},\phantom{bbbbb}Y(d_R)=-\tfrac{2}{3}.
\end{eqnarray*}

We have demonstrated in detail elsewhere \cite{hosek} that the
strong exchanges of the dynamically massive flavor gluons indeed
play the role of the extended Higgs sector of the Yanagida model.
Here we emphasize their selective dynamical role which results in
the computable neutrino mass spectrum in the seesaw form.

\section{II. Fermion mass spectrum}
Our task is to generate the fermion proper self-energy $\Sigma(p)$
in the full fermion propagator
\begin{equation}
S^{-1}(p)=\slashed p - \Sigma(p)
\end{equation}
dynamically by the strong flavor gluon interactions of $SU(3)_f$. In
general $\Sigma(p)$ is a complex $3 \times 3$ matrix, the solution
of the Schwinger-Dyson equation \cite{pagels}
\begin{widetext}
\begin{eqnarray}
\Sigma(p)=3\int \frac{d^4k}{(2\pi)^4}\frac{\bar
h^2_{ab}((p-k)^2)}{(p-k)^2}T_a(R)\Sigma(k)[k^2-\Sigma^{+}(k)\Sigma(k)]^{-1}T_b(L)
\label{Sigma}
\end{eqnarray}
\end{widetext}
depicted in Fig.1.

To make the formidable task of the dynamical fermion mass generation
tractable we resort to approximations. First, without loss of
generality we fix the external (Euclidean) momentum as $p=(p,\vec
0)$, and integrate over angles. Integration over the momenta only up
to $\Lambda$ means that the resulting model is not asymptotically,
but strictly free above this scale:
\begin{equation}
\Sigma(p)=\int_0^{\Lambda}k^3dk
K_{ab}(p,k)T_a(R)\Sigma(k)[k^2+\Sigma^{+}\Sigma]^{-1}T_b(L).
\label{Sigmasep}
\end{equation}
Because of the unknown behavior of $\bar h_{ab}^2$ below $\Lambda$
the kernel $K_{ab}(p,k)$, separately symmetric in momenta and flavor
octet indices, is {\it entirely unknown}.
\begin{figure}[ht]
\includegraphics[width=0.45\textwidth] {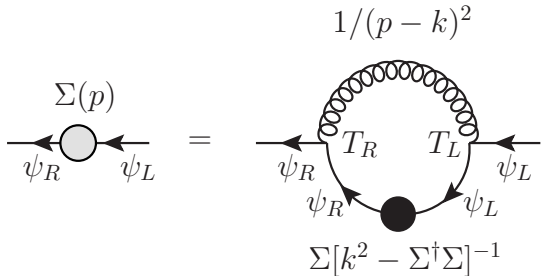}
\caption[99]{At high momenta the  coupling matrix $\bar h_{ab}^2$ is
small and known (asymptotic freedom). Below $\Lambda$ the coupling
matrix $\bar h_{ab}^2$ is large and entirely unknown.}
\label{obrazek}
\end{figure}
Second and most important, for this unknown kernel we make a
separable Ansatz
\begin{equation}
K_{ab}(p,k)=\frac{3}{4\pi^2} \frac{g_{ab}}{pk}.\label{sep}
\end{equation}
It is conceptually quite similar to the separable Ansatz  for the
potential in the BCS superconductor which leads to the explicit
solution of the BCS gap equation \cite{fetter}. Why the simple
separable approximation taking into account only the opposite
momenta around Fermi surface is phenomenologically so successful was
a mystery for years. It was theoretically clarified much later by
Polchinski \cite{polchinski}.

Here $g_{ab}$ are the effective dimensionless low-energy parameters
which characterize the {\it complete spontaneous breakdown of the
gauge $SU(3)_f$}. Very important question is how many independent
parameters are necessary for such a breakdown. We believe the answer
is {\it three}. We present two independent explicit illustrations of
this statement.

First illustration is provided by the Higgs mechanism for $SU(3)_f$
with the condensing scalar sextet $\chi$, the complex symmetric $3
\times 3$ matrix \cite{yanagida,hosek}. Here the spontaneous gauge
boson and fermion mass generations are two generically independent
phenomena: (i) Eight different gauge boson masses squared are
expressed in terms of three parameters $(v_1,v_2,v_3)$, the diagonal
vacuum expectation values of $\chi$:
\begin{widetext}
\begin{multline*}
M^2 \sim \left[
\begin{array}{cccccccc}
(v_1+v_2)^2 & 0 & 0 & 0 & 0 & 0 & 0 & 0\\
0 & (v_1-v_2)^2 & 0 & 0 & 0 & 0 & 0 & 0\\
0 & 0 & 2(v_1^2+v_2^2) & 0 & 0 & 0 & 0 & \frac{2}{\sqrt3}(v_1^2-v_2^2)\\
0 & 0 & 0 & (v_1+v_3)^2 & 0 & 0 & 0 & 0\\
0 & 0 & 0 & 0 & (v_1-v_3)^2 & 0 & 0 & 0\\
0 & 0 & 0 & 0 & 0 & (v_2+v_3)^2 & 0 & 0\\
0 & 0 & 0 & 0 & 0 & 0 & (v_2-v_3)^2 & 0\\
0 & 0 & \frac{2}{\sqrt3}(v_1^2-v_2^2) & 0 & 0 & 0 & 0 &
\frac23(v_1^2+v_2^2+4v_3^2)
\end{array}
\right]
\end{multline*}
\end{widetext}
(with diagonalization of the $(3,8)$ block understood). (ii) The
$SU(3)_f$ invariant Yukawa interaction with the sextet $\chi$ gives
rise to three different Majorana masses to chiral neutrinos
proportional to $v_i$ provided they transform as a triplet of
$SU(3)_f$.

Second illustration is provided by spontaneous breakdown of the
gauge $SU(3)_f$ by the dynamically generated Majorana masses of
neutrinos transforming as the triplet \cite{hosek}. Here the
phenomena of spontaneous generation of the gauge and the fermion
masses are self-consistently related. Three different Majorana
masses are found by solving the SD equation, and eight gauge boson
masses are the functions of these three Majorana masses computed by
the Pagels-Stokar formula.

We conclude that the matrix $g_{ab}$ in the present
'semi-microscopic' approach should be expressible as a function of
three dimensionless parameters. The matrix $g \sim M^2/\Lambda^2$ is
the simplest weak-coupling prototype of such a relation, with
positive-definiteness relaxed. Ultimately, however, because the
group $SU(3)$ is simple, even these three parameters should be
calculable.

In separable approximation the SD equation is immediately formally
solved:
\begin{equation}
\Sigma(p)=\frac{\Lambda^2}{p}T_a(R)\Gamma_{ab}T_b(L)\equiv\frac{\Lambda^2}{p}\sigma
\label{sol}
\end{equation}
The difficult part is that the numerical matrix $\Gamma$ has to
fulfil the homogeneous nonlinear algebraic self-consistency
condition (gap equation).

The obtained momentum dependence of $\Sigma(p)\sim 1/p$ is not
without support. Because the masses of flavor gluons come out huge
it is justified to think heuristically of the dynamically generated
fermion masses in terms of the four-fermion interaction of Nambu and
Jona-Lasinio \cite{njl}. In a series of papers \cite{mannheim}
Philip Mannheim argues that the theoretically  consistent treatment
of the fermion mass generation by the four-fermion dynamics should
result in $\Sigma(p) \sim 1/p$.

\subsection{1. Fermion masses with flavor mixing}
{\bf For the right-handed neutrinos} the effective Majorana mass
term has the form
\begin{equation}
{\cal L}_{Majorana}=-\tfrac{1}{2}(\bar
\nu_{R}\Sigma_M(p)(\nu_{R})^{{\cal C}}+ h.c.)\label{majorana}
\end{equation}

In flavor space it therefore transforms as
\begin{equation}
\bar 3 \times \bar 3 = 3_a + \bar 6_s
\end{equation}
where the subscripts abbreviate the antisymmetric (a) and symmetric
(s) representations.

Because of the Pauli principle $\sigma_M$ in (\ref{sol}) is a
general complex symmetric $3 \times 3$ matrix of the sextet. It can
be put into a positive-definite real diagonal matrix $\gamma^M$ by a
constant transformation
\begin{equation}
\sigma_M=U^{+}\gamma^M U^{*}.
\end{equation}
Then, the gap equation becomes
\begin{equation}
\gamma^M=UT_a(R)U^{+}g_{ab}I(\gamma^M)U^{*}T_b(L)U^{T}
\label{gammaM}
\end{equation}
where
\begin{eqnarray}
I(\gamma)&=&\frac{3}{16\pi^2}\gamma
\int_{0}^{1}\frac{dx}{x+\gamma^2}= \frac{3}{16\pi^2}\gamma \rm
ln\frac{1+\gamma^2}{\gamma^2}.\nonumber\\
\end{eqnarray}

The diagonal entries of the equation (\ref{gammaM}) determine the
sterile neutrino masses, the nondiagonal entries provide relations
for the mixing angles and the new CP-violating phases. These phases
are most welcome as a source of an extra CP violation needed for
understanding of the baryon asymmetry of the Universe
\cite{leptogenesis}. Because $\nu_R$ transforms as a triplet,
$T_a(R)=\tfrac{1}{2}\lambda_a$. The charge-conjugate $\nu_R$ is a
left-handed field i.e., $(\nu_{R})^{{\cal C}}=(\nu)^{{\cal C}}_L$.
Because charge conjugation is essentially the hermitian conjugation,
$T_b(L)=-\tfrac{1}{2}\lambda^{*}_b=-\tfrac{1}{2}\lambda^{T}_b$ are
the generators of antitriplet.

{\bf For Dirac fermions} the effective Dirac term has the form
\begin{equation}
{\cal L}_{Dirac}=-(\bar f_{R}\Sigma_D(p)f_L+ h.c.)\label{dirac}
\end{equation}
In flavor space it therefore transforms as
\begin{equation}
\bar 3 \times 3 = 1 + 8.
\end{equation}
The general complex $3 \times 3$ matrix $\sigma_D$ can be put into a
positive-definite real diagonal matrix $\gamma^D$ by a constant
bi-unitary transformation:
\begin{equation}
\sigma_D=U^{+}\gamma^D V.
\end{equation}
The gap equation becomes
\begin{equation}
\gamma^D=UT_a(R)U^{+}g_{ab}I(\gamma^D)VT_b(L)V^{+}. \label{gammaD}
\end{equation}
The diagonal entries of the equation (\ref{gammaD}) determine the
fermion masses, the nondiagonal entries provide relations for the
CKM mixing angles and the SM CP-violating phase. Because all chiral
fermion fields of the model transform as triplets,
$T_a(R)=\tfrac{1}{2}\lambda_a$ and $T_b(L)=\tfrac{1}{2}\lambda_b$.

\vspace{3mm}

\subsection{2. Majorana masses $M_M$ and Dirac masses $m_D$}
In the following we set the fermion mixing matrices to the unit
matrix and show that the Majorana masses $M_M$ of the right-handed
neutrinos come out naturally huge of order $\Lambda$, whereas the
Dirac masses of the fermions of the model are naturally small
compared to $\Lambda$. 'Naturally' means that in the Lagrangian
there are no parameters (effective couplings $g_{ab}$) which would
differ by many orders of magnitude. Huge mass ratios emerge only
{\it in solutions} of the underlying field equations.

Without mixing the Majorana and Dirac gap equations are \cite{hosek}
\begin{equation}
\gamma^M=-\tfrac{1}{4}\lambda_a g_{ab}
I(\gamma^M)\lambda_b^{T}\label{A2}
\end{equation}
and
\begin{equation}
\gamma^D=\tfrac{1}{4}\lambda_a g_{ab} I(\gamma^D)\lambda_b
\label{A1},
\end{equation}
respectively. Because $\Sigma(p)\equiv \tfrac{\Lambda^2}{p}\gamma$,
the fermion mass, both Dirac and Majorana, defined as a pole of the
full fermion propagator is
\begin{equation*}
m \equiv \Sigma(p^2=m^2)=\Lambda \gamma^{1/2}
\end{equation*}

With $g_{11}, g_{22}, g_{33}, g_{38}, g_{44}, g_{55}, g_{66},
g_{77}, g_{88}$ different from zero the right hand sides of
equations (\ref{A1}) and (\ref{A2}) are the diagonal matrices. The
equations themselves can be rewritten as
\begin{equation}\label{A3}
\gamma_i^{D/M}=\sum_{k=1}^3 \alpha_{ik}^{D/M}\gamma_k^{D/M}
\ln\frac{1+(\gamma_k^{D/M})^2}{(\gamma_k^{D/M})^2}
\end{equation}
where
\begin{widetext}
\begin{equation}\label{alpha}
\alpha^{D/M}=\frac{3}{64\pi^2}\left(\begin{array}{ccc}
\pm\left(g_{33}+\frac{2}{\sqrt{3}}g_{38}+\frac{1}{3}g_{88}\right) & g_{22}\pm g_{11} & g_{55}\pm g_{44}\\
g_{22}\pm g_{11} & \pm\left(g_{33}-\frac{2}{\sqrt{3}}g_{38}+\frac{1}{3}g_{88}\right) & g_{77}\pm g_{66}\\
g_{55}\pm g_{44} & g_{77}\pm g_{66} & \pm \frac{4}{3}g_{88}
\end{array}
\right)
\end{equation}
\end{widetext}
and the upper and lower signs correspond to the Dirac fermion masses
and the Majorana neutrino masses, respectively.

The form of the matrix $\alpha$ suggests simplifications. In the
following we take only
\begin{equation*}
g_{33},g_{38},g_{88};g_{11}=-g_{22},g_{44}=-g_{55},g_{66}=-g_{77}
\end{equation*}
different from zero (and expect that they are not independent).

(A) The matrix gap equation for the Dirac masses $m_{iD}$ becomes
diagonal and decoupled, and it is easily solved. Provided the
combinations
\begin{eqnarray*}
\alpha_1&=&\tfrac{3}{64\pi^2}(g_{33}+\tfrac{2}{\sqrt
3}g_{38}+\tfrac{1}{3}g_{88})\\
\alpha_2&=&\tfrac{3}{64\pi^2}(g_{33}-\tfrac{2}{\sqrt
3}g_{38}+\tfrac{1}{3}g_{88})\\
\alpha_3&=&\tfrac{3}{64\pi^2}\tfrac{4}{3}g_{88}
\end{eqnarray*}
are all positive and all $\alpha_i \ll 1$, the resulting universal
flavor-splitting Dirac mass formula is
\begin{equation}
\boxed{m_{iD}=\Lambda \phantom{b} \rm exp \phantom{b}
(-1/4\alpha_i)}.\label{mD}
\end{equation}

\vspace{3mm}

(B) Finding the solution for the Majorana masses is less
straightforward and we have to resort to simple numerical
demonstration. First, for $g_{11}=g_{44}=g_{66}=0$, the gap
equations for the Majorana masses have no solution because of the
minus sign in front of the $\alpha_i$. Consequently, $(g_{11},
g_{44}, g_{66})\neq 0$. Second, in the case of sterile Majorana
neutrinos we are not aware of the necessity of the hierarchical mass
spectrum. With the constants $\alpha_i$ fixed by the numerical
values of the Dirac masses the equations (\ref{A3}) for $\gamma_i^M$
can be viewed as a system of three inhomogeneous linear equations
for the unknown $(g_{11},g_{44},g_{66})$:
\begin{widetext}
$$
-\frac{1}{2}\left(\begin{array}{ccc}
I(\gamma_2^M)&I(\gamma_3^M)&0\\
I(\gamma_1^M)&0& I(\gamma_3^M)\\
0&I(\gamma_1^M)&I(\gamma_2^M)
\end{array}\right)
\left(\begin{array}{c} g_{11} \\g_{44} \\g_{66}
\end{array}\right)=
\left(\begin{array}{c}
\gamma_1^M+\tfrac{16\pi^2}{3}\alpha^{D}_{11} I(\gamma_1^M)\\
\gamma_2^M+\tfrac{16\pi^2}{3}\alpha^{D}_{22} I(\gamma_2^M)\\
\gamma_3^M+\tfrac{16\pi^2}{3}\alpha^{D}_{33} I(\gamma_3^M)
\end{array}\right).
$$
\end{widetext}
This set of equations has a solution for any set of $\gamma_i^M >
0$.

\newpage

For illustration that
\begin{equation}
\boxed{M_{iM} \sim \Lambda}
\end{equation}

we put $(\gamma_1^M,\gamma_2^M,\gamma_3^M)=(0.1,0.2,0.3)$ and
$(\gamma_1^D,\gamma_2^D,\gamma_3^D)=(10^{-20},10^{-22},10^{-26})$.
This corresponds approximately to the hierarchy for charged leptons
provided $\Lambda=10^{10}\,$GeV. Then
\begin{widetext}
$$g=\left(
\begin{array}{cccccccc}
 8.08101 & 0 & 0 & 0 & 0 & 0 & 0 & 0 \\
 0 & -8.08101 & 0 & 0 & 0 & 0 & 0 & 0 \\
 0 & 0 & 1.7425 & 0 & 0 & 0 & 0 & 0.0899893 \\
 0 & 0 & 0 & -21.8124 & 0 & 0 & 0 & 0 \\
 0 & 0 & 0 & 0 & 21.8124 & 0 & 0 & 0 \\
 0 & 0 & 0 & 0 & 0 & -34.029 & 0 & 0 \\
 0 & 0 & 0 & 0 & 0 & 0 & 34.029 & 0 \\
 0 & 0 & 0.0899893 & 0 & 0 & 0 & 0 & 1.31887 \\
\end{array}
\right)$$
\end{widetext}
It is important that the precise size and hierarchy of $\gamma_i^D$
does not play any important role for the numerical values of
$\gamma_i^M$.

It is easy to understand qualitatively the enormous, yet natural
difference between the huge Majorana masses $M_R$ of the
right-handed neutrinos and the tiny Dirac masses $m_D$ of leptons
and quarks: The difference between $\bar 3 \times \bar 3$ and $\bar
3 \times 3$ translates into different combinations of parameters in
$g_{ab}$ which determine $M_R$ and $m_D$, respectively.

\subsection{3. Why not $m_M$ ?}
The dynamically generated fermion masses in the chiral $SU(3)_f
\times SU(2)_L \times U(1)_Y$ gauge model imply spontaneous
breakdown of this symmetry down to the unbroken $U(1)_{em}$.
Consequently, there must be exactly $8 + 3$ 'would-be' NG bosons.
With the weakly coupled Higgs sector in the Yanagida model they are
pre-prepared in the elementary scalar Higgs fields
$\chi^{ab}(6,1,0)$, $\phi^a(8,2,1)$ and $\phi^0(1,2,1)$. We have
confirmed in \cite{hosek} by analyzing the {\it symmetry structure}
of the NG poles in Ward-Takahashi identities that these 'would-be'
NG bosons are in fact parts of the fermion-antifermion composites
\begin{eqnarray*}
\chi^{ab}(6,1,0) & \sim & \nu_R^a (\bar \nu_R)^{{\cal C}b}\\
\phi^a(8,2,1)    & \sim & (\bar d_R \lambda^a q_L + \bar e_R \lambda^a l_L )\\
\phi^0(1,2,1)    & \sim  &(\bar d_R q_L + \bar e_R l_L )\\
\end{eqnarray*}
formed by strong exchanges of dynamically massive flavor gluons. The
remaining components show up in the spectrum as massive composite
Higgs-like particles.

Why not the elementary Higgs field $\phi(6,3,-2)$ \cite{seesaw2}?
Simply because such a decision is solely in hands of the model
builder.

In model with the Higgs sector replaced by strong exchanges of
dynamically massive flavor gluons the situation is entirely
different. If allowed these exchanges would obviously generate
$m_M=M_M$ which would entirely ruin the seesaw construction. We are
obliged to {\it assume} the dynamical generation of $m_M$ and check,
if possible, whether such an assumption is theoretically consistent.
Symmetry considerations themselves cannot replace the issues of
dynamics.

(i) The assumption implies specific pattern of spontaneous breakdown
of $SU(3)_f \times SU(2)_L \times U(1)_Y$ symmetry.

(ii) Pauli principle and the symmetry structure of the NG poles in
the WT identity suggest that the NG bosons are the composites of the
lepton doublet and its charge conjugate
\begin{equation}
\phi^{ab}(6,3,-2) \sim (\bar l_L)^{{\cal C}a}i\tau_2 \vec \tau l_L^b
\end{equation}
carrying the electric charges $(0, +, ++)$.

(iii) Strong Coulomb repulsion in the doubly charged component {\it
without confining force} suggests that the composite $\phi(6,3,-2)$
should not be formed. Consequently, neither the condensate of its
neutral component, $m_M$, should be dynamically generated.

(iv) It is mandatory to check whether the dynamical argument applies
also to other composite NG bosons. In $\chi$ all constituents are
electrically neutral and there is no problem. In singly charged
components of both $\phi^a$ and $\phi$ made of a neutrino and a
charged lepton there is no problem. In their singly charged quark
components there is a strong Coulomb repulsion in ($\bar d u$). It
is, however, safely overwhelmed by the QCD confining force as in
ordinary electrically charged hadrons.

\section{III. Conclusion}
Not surprisingly, the obtained flavor splitting of the fermion
masses (\ref{mD}) is universal for all types of fermions. With
electroweak gauge interactions not actively involved there is
nothing in the model which would distinguish between fermions with
different weak hypercharges or electric charges i.e., in given
generation all Dirac masses must come out equal. This is,
nevertheless, enough to break down spontaneously the gauge symmetry
$SU(3)_f \times SU(2)_L \times U(1)_Y$ down to $U(1)_{em}$. The
electroweak gauge interactions, treated as weak external
perturbations, thus become short-range: $W$ and $Z$ absorb the
underlying composite multi-component 'would-be' NG bosons as their
longitudinal polarization states and the Weinberg relation
$m_W/m_Z=\rm cos \phantom{b} \theta_W$ is exact.

Since the electroweak gauge interactions remain weak up to the
Planck scale they themselves cannot generate the fermion masses
dynamically as suggested long ago \cite{mahan}. They do, however,
provide {\it electroweak radiative corrections} to the kernel of the
SD equation which distinguish the masses $m_{il}$ of charged leptons
and masses $m_{iu}$ and $m_{id}$ of quarks with charges $2/3$ and
$-1/3$, respectively. At short distances there is also the universal
positive QCD radiative correction which makes quarks in given
generation heavier than the corresponding leptons. It is important
that these corrections do not contain any new free parameters
\cite{hosek1}. We believe that they should be greatly amplified by
the exponentials in the universal formula (\ref{mD}). How to
incorporate them into the separable Ansatz in the SD equation
remains to be demonstrated.

We do know, however, that the radiative corrections mentioned above
vanish for the Dirac neutrino masses: $Y(\nu_R)=0$ and there are no
QCD corrections. Therefore, within the given rules of the game the
computation of $m_{iD}$ is the exact computation of the {\it Dirac
neutrino mass matrix}. Because of sterility there are no electroweak
corrections to $M_M$ either. Hence the computation of the whole
neutrino seesaw mass matrix by strong flavor gluon exchanges treated
in a separable approximation is 'exact'.

I am grateful to Philip Mannheim for interest in the model, and to
Michal Malinsk\' y for valuable comments.

\end{document}